\documentstyle{article}
\begin{document}
\sloppy

\begin{center}
{\Large \bf
Dynamical Probability Distribution Function of\\
the SK Model at High Temperatures
}
\end{center}

\vspace{1cm}

\begin{center}
Hidetoshi Nishimori and Michiko Yamana
\end{center}

\begin{center}
Department of Physics, Tokyo Institute of Technology,
 \\Oh-okayama, Meguro-ku, Tokyo 152, Japan
\end{center}
\vspace{5mm}

\begin{abstract}
The microscopic probability distribution function of the
Sherrington-Kirkpatrick (SK) model of spin glasses is calculated explicitly
as a function of time by a high-temperature expansion.
The resulting formula
to the third order of the inverse temperature shows that an
assumption made by
Coolen, Laughton and Sherrington in their recent theory of
dynamics is violated.
Deviations of their theory from exact results are estimated quantitatively.
Our formula also yields explicit expressions of the time dependence
of various macroscopic physical quantities when the temperature is suddenly
changed within the high-temperature region.
\end{abstract}
\vspace{1cm}

The dynamics of the Sherrington-Kirkpatrick (SK) model
of spin glasses~\cite{ref:SK}
has been discussed for many years.~\cite{ref:Binder-Young} However, no
explicit exact expressions of evolution equations have been given
for macroscopic physical quantities such as the
magnetization or susceptibility.

For a closely related problem, the dynamics of the Hopfield model of neural
networks,~\cite{ref:Hopfield} the situation is somewhat different.
For a finite number of embedded patterns $p$, it is possible to
write down a set of evolution  equations of macroscopic
order parameters explicitly.
\cite{ref:CR,ref:SNO,ref:Hemmen} If, on the other hand,
the number of patterns
$p$ is
proportional to the system size $N$, the problem is quite nontrivial and
there has been no exact theory describing the time dependence of physical
quantities.

Coolen and Sherrington (CS)~\cite{ref:CS1,ref:CS2} claimed to have
derived the exact solution of this
latter problem (in which $p$ is proportional to $N$)
using physical assumptions on the microscopic
probability distribution function.
They assumed that the microscopic probability distribution function
is a constant within a subspace in which a limited number of
macroscopic order parameters take fixed values.
They called this property equipartitioning. In the equilibrium limit
the equipartitioning property holds trivially because the Boltzmann factor
is a constant for a given energy.   In general nonequilibrium
situations, however, it is not clear whether or not equipartitioning
is a valid assumption.
Ozeki and Nishimori~\cite{ref:ON} showed by Monte
Carlo simulations that numerical data for the noise
distribution function, in terms
of which the time evolution of macroscopic quantities is
determined, deviate
from the predictions of CS.  Thus the method of CS
yields approximations, not exact results, for the dynamics
of the Hopfield model.
CS then applied the same idea to the SK model,~\cite{ref:CS3} and discussed
the dynamics in terms of approximate closed-form evolution
equations of a few macroscopic order parameters.

The most recent development along this line is due to Coolen, Laughton
and Sherrington (CLS).~\cite{ref:CLS,ref:LCS}
This recent work is a sophistication
of their previous method, using a continuous function instead
of a few order
parameters to describe the dynamics of the macroscopic state of the system.
The basic assumption of equipartitioning of the microscopic probability
distribution function
remains essentially intact.
More precisely, they assumed that the microscopic probability
distribution function $p_t({\bf \sigma})$
is a constant ({\it i.e.}, it does not depend on the spin configuration
${\bf \sigma}$) once the value of the single-site spin-field
distribution function
\begin{equation}
  D(\varsigma, h)=\frac{1}{N}\sum_i \delta_{\varsigma,\sigma_i}
   \delta(h-h_i)
   \label{spin-field}
\end{equation}
is given, where the local field is
\begin{equation}
  h_i({\bf \sigma})=\sum_{j(\ne i)} J_{ij}\sigma_j .
  \label{local-field}
\end{equation}
They conjectured, based partly  on comparison with
numerical data, that this sophisticated version is either exact
or a very good approximation.

In the present letter we solve the master equation of the
SK model explicitly by
a high-temperature expansion.
The resulting formula for the microscopic probability
distribution function shows
that the assumption of equipartitioning is violated.
Thus the CLS theory is not exact
even in its sophisticated form.
We derive quantitative estimations of deviations of their theory from exact
results.  We also show that
our formula is useful for
predicting the behavior of physical quantities after
a sudden change of temperature
within the high-temperature region.

The SK model is described by $N$ Ising spins interacting with each other
via random infinite-range interactions $J_{ij}$:
\begin{equation}
   H({\bf \sigma}) =-\sum_{i<j}J_{ij} \sigma_i\sigma_j .
   \label{Hamiltonian}
\end{equation}
$J_{ij}$ represents quenched disorder, the values of which are
obtained independently from
a Gaussian distribution with mean $J_0/N$ and variance $J^2/N$.  The
microscopic probability distribution function $p_t({\bf \sigma})$
obeys the master equation
\begin{equation}
   \frac{1}{p_t({\bf \sigma})}\frac{{\rm d}}{{\rm d}t}p_t({\bf \sigma})
   =\frac{1}{p_t({\bf \sigma})}\sum_k
     {p_t(F_k {\bf \sigma})}w_k(F_k{\bf \sigma})
     -\sum_kw_k({\bf \sigma}).
   \label{master}
\end{equation}
Here $F_k$ is a single spin flip operator
\[
F_k\Phi ({\bf \sigma})
\equiv \Phi (\sigma_1,\cdots,-\sigma_k,\cdots,\sigma_N)
\]
and the transition rate is defined by
\begin{equation}
 w_k({\bf \sigma})=\frac{1}{2}\left\{
    1-\sigma_k\tanh \beta h_k({\bf \sigma})\right\} .
  \label{t-rate}
\end{equation}
The inverse temperature is denoted as $\beta$.

Let us solve the master equation (\ref{master}) by a high-temperature
expansion in the form
\begin{equation}
 p_t({\bf \sigma})=\exp \left\{ \beta f_t({\bf \sigma})
   +\beta^2 g_t({\bf \sigma})+\beta^3 u_t({\bf \sigma})+\cdots\right\}
   \label{expansion}
\end{equation}
under the initial condition
\begin{equation}
   p_{t=0}({\bf \sigma})=
   \exp \{-\beta_0H({\bf \sigma})\}.
   \label{initial-distr}
\end{equation}
Normalization of the distribution function (\ref{expansion})
is irrelevant for the
following argument.
Inserting eq.~(\ref{expansion}) into the master equation (\ref{master})
and expanding the result in powers of $\beta$, we obtain the right-hand side
of the master equation as
\begin{eqnarray}
   & & \frac{1}{p_t({\bf \sigma})}\sum_k
    {p_t(F_k {\bf \sigma})}w_k(F_k{\bf \sigma})
    -\sum_k w_k({\bf \sigma})
   \nonumber
   \\
   &\sim&\sum_k\left\{ 1+\beta\Delta_k f_t+\beta^2
   \left( \frac{1}{2} (\Delta_k f_t)^2
      +\Delta_k g_t \right) \right.
    \nonumber
    \\
   & & {} +\left. \beta^3 \left( \frac{1}{6} (\Delta_k f_t)^3
   +\Delta_k f_t
      \Delta_k g_t +\Delta_k u_t \right) +\cdots\right\}
\frac{1}{2} \left( 1+\beta\sigma_k h_k
      -\frac{1}{3} \beta^3 \sigma_k h_k^3+\cdots\right)
    \nonumber
    \\
   & & {} -\frac{1}{2} \sum_k \left( 1-\beta\sigma_k h_k +
   \frac{1}{3} \beta^3
      \sigma_k h_k^3-\cdots\right)
    \label{high-expand}
\end{eqnarray}
where $\Delta_k f_t\equiv f_t(F_k{\bf \sigma})-f_t({\bf \sigma})$,
and similarly for $\Delta_k g_t$ and $\Delta_k u_t$.

To evaluate the first-order term $f_t({\bf \sigma})$
in eq. (\ref{expansion}),
we retain only the terms proportional to $\beta$ in eq.~(\ref{high-expand})
and compare these with the left-hand side of eq.~(\ref{master}).
The result is
\begin{equation}
  \frac{{\rm d}f_t}{{\rm d}t}=\frac{1}{2}\sum_k\Delta_k f_t -
  2H({\bf \sigma}) .
  \label{df-dt}
 \end{equation}
The following form is a possible solution of this equation:
\begin{equation}
  f_t({\bf \sigma}) = a(t)H({\bf \sigma}) .
  \label{f-general}
\end{equation}
Inserting eq.~(\ref{f-general}) into eq.~(\ref{df-dt}), we find
\[
  \dot{a} (t) = -2a(t)-2 .
  \]
This equation is easily solved under the initial condition $a(t=0)=
-\beta_0/\beta$, which corresponds to eq.~(\ref{initial-distr}), as
\begin{equation}
  a(t)= \left( 1-\frac{\beta_0}{\beta}\right) e^{-2t} -1 .
  \label{a(t)}
\end{equation}
Since eq.~(\ref{df-dt}) is a first-order differential equation, the above
expression (\ref{f-general}) with eq.~(\ref{a(t)})
represents the unique solution.
The first-order contribution has thus been obtained as
\begin{equation}
  p_t({\bf \sigma})=\exp \big[ \{ (\beta -\beta_0 )e^{-2t} -\beta \}
   H({\bf \sigma}) \big] .
   \label{1st}
\end{equation}
It is noted here that $\beta_0$ should be at most of the same order
as $\beta$ because $a(t)$ in eq.~(\ref{f-general})
has implicitly been assumed to be of order unity.

Similarly, from the second-order term of $\beta$
in eq.~(\ref{high-expand}), we have
\begin{equation}
  \frac{{\rm d}g_t}{{\rm d}t} = \frac{1}{2}\sum_k\Delta_k g_t
  + (\alpha^2 e^{-4t} -\alpha e^{-2t}) \sum_k h_k^2 ,
  \label{dg-dt}
\end{equation}
where $\alpha\equiv 1-\beta_0/\beta$.  This equation
suggests a solution of the following form:
\begin{equation}
  g_t({\bf \sigma}) = b_1 (t)\sum_i h_i^2 +b_2 (t).
  \label{g-general}
\end{equation}
Inserting eq.~(\ref{g-general}) into eq.~(\ref{dg-dt})
and using the relation
\[
  \Delta_k (h_i)^2 = -4h_i J_{ik}\sigma_k + 4 J_{ik}^2 ,
  \]
we find
\begin{equation}
  \dot{b_1} (t) = -2b_1 (t) +\alpha^2 e^{-4t} -\alpha e^{-2t}.
  \label{dot-b1}
\end{equation}
Since the $b_2(t)$ term in eq.~(\ref{g-general}) does not depend on
the spin configuration and affects only the overall normalization of the
probability distribution function (\ref{expansion}), we do not
write down  the evolution equation for $b_2(t)$ here.
The solution of
eq.~(\ref{dot-b1}) is
\begin{equation}
  b_1 (t)= -\frac{\alpha^2}{2} e^{-4t}
  -\left(\alpha t-\frac{\alpha^2}{2}\right) e^{-2t}.
  \label{b_1 (t)}
\end{equation}
The initial condition is $b_1(t=0)=0$ because the first-order
contribution, eq.~(\ref{1st}),
already satisfies the initial condition (\ref{initial-distr}).

The third-order term of $\beta$ in eq.~(\ref{high-expand}) yields
\begin{eqnarray}
 \frac{{\rm d}u_t}{{\rm d}t} &=& \frac{1}{2}\sum_k\Delta_k u_t
 -4J^2 (2\alpha e^{-2t} -1) b_1 (t) H({\bf \sigma})
 \nonumber
 \\
 & & {} +(\frac{2}{3} \alpha^3 e^{-6t} -\alpha^2 e^{-4t})
 \sum_k\sigma_k h_k^3
 -2(2\alpha e^{-2t} -1) b_1 (t)\sum_{k,l} J_{kl} h_k h_l.
 \label{du-dt}
\end{eqnarray}
The following form seems an appropriate solution of this equation:
\begin{equation}
  u_t({\bf \sigma}) = c_1 (t)H({\bf \sigma})
  +c_2 (t)\sum_i\sigma_i h_i^3
  +c_3 (t)\sum_{i,j} J_{ij} h_i h_j.
 \label{u-general}
\end{equation}
The first term of the right-hand side of eq.~(\ref{du-dt}) is calculated
under the assumption (\ref{u-general}) as
\begin{eqnarray}
 &\frac{1}{2}& \sum_k \Delta_k u_t = c_1 (t)
 \left\{-2 H({\bf \sigma})\right\}
 \nonumber
 \\
 & & {} + c_2 (t)\left\{-4\sum_k\sigma_k h_k^3 -12 J^2 H({\bf \sigma})
                        -4\sum_{k,l} J_{kl}^3 \sigma_k \sigma_l \right\}
 \nonumber
 \\
 & & {} + c_3 (t)\left\{-2\sum_{k,l} J_{kl} h_k h_l
                        +2\sum_{k,l,m} J_{kl} J_{km} J_{lm}\right\}.
\end{eqnarray}
The expressions including $J_{kl}^3$ or $J_{kl} J_{km} J_{lm}$ do
not appear in eq.~(\ref{u-general}).
Thus eq.~(\ref{u-general}) may at first sight
appear inappropriate as the solution
of eq.~(\ref{du-dt}).
However, a simple order estimate using the relation
\[
  J_{kl} = \frac{J_0}{N} + \frac{J z_{kl}}{\sqrt{N}},
  \]
where $z_{kl}$ is a Gaussian variable with vanishing mean and unit variance,
reveals that these $J^3$-terms are at most of order $\sqrt{N}$.
Therefore these $J^3$-contributions can be ignored compared to the
other terms which are all of order $N$.  In this way, we find that
eq.~(\ref{u-general}) gives a consistent solution of eq.~(\ref{du-dt})
with coefficients satisfying
\begin{eqnarray}
  \dot{c_1} (t) &=& -2c_1 (t) -12J^2 c_2 (t)
  \nonumber
  \\
  & & {} -4J^2 b_1 (t) (2\alpha e^{-2t} -1)
  \nonumber
  \\
  \dot{c_2} (t) &=& -4c_2 (t) + \frac{2}{3}\alpha^3 e^{-6t}
  -\alpha^2 e^{-4t}
  \nonumber
  \\
  \dot{c_3} (t) &=& -2c_3 (t) -2b_1 (t) (2\alpha e^{-2t} -1).
  \nonumber
\end{eqnarray}
These differential equations are solved as
{
\setcounter{enumi}{\value{equation}}
\addtocounter{enumi}{1}
\setcounter{equation}{0}
\renewcommand{\theequation}{\theenumi\alph{equation}}
\begin{eqnarray}
  c_1 (t) &=& J^2\left\{-2\alpha^3 e^{-6t}
                      + (4\alpha^3 -4\alpha^2 -10\alpha^2 t) e^{-4t} \right.
  \nonumber
  \\
  & & {} + \left.
        (2\alpha^2 t -2\alpha t^2 +4\alpha^2 -2\alpha^3) e^{-2t}\right\}
  \label{c_1 (t)}
  \\
  c_2 (t) &=& -\frac{1}{3}\alpha^3 (e^{-6t} - e^{-4t}) -\alpha^2 t e^{-4t}
  \label{c_2 (t)}
  \\
  c_3 (t) &=& -\frac{1}{2}\alpha^3 e^{-6t}
            +(\alpha^3 -\frac{1}{2}\alpha^2 -2\alpha^2 t) e^{-4t}
  \nonumber
  \\
  & & {} + (\alpha^2 t -\alpha t^2 -\frac{1}{2}\alpha^3
  +\frac{1}{2}\alpha^2)
       e^{-2t}
  \label{c_3 (t)}
\end{eqnarray}
\setcounter{equation}{\value{enumi}}
}
under the initial condition $c_1=c_2=c_3=0$.

We have obtained the probability distribution function of the
SK model at high temperature as
\begin{eqnarray}
   p_t({\bf \sigma}) = &\exp& \left[ \beta a(t) H({\bf \sigma})
            + \beta^2 b_1 (t) \sum_i h_i^2 \right.
  \nonumber
  \\
  & & {} + \beta^3 \left\{ c_1(t) H({\bf \sigma}) +
         c_2 (t) \sum_i \sigma_i h_i^3 \right.
+ \left. \left. c_3 (t) \sum_{i,j} J_{ij} h_i h_j \right\} +
         \cdots \right] ,
   \label{result}
\end{eqnarray}
where $a (t), b_1 (t), c_1 (t), c_2 (t)$ and $c_3 (t)$ are given in
eqs.~(\ref{a(t)}), (\ref{b_1 (t)}) and (\ref{c_1 (t)}) - (\ref{c_3 (t)}).
The result (\ref{result}) shows that there exists
a term proportional to $\sum J_{ij}h_{i}h_{j}$ representing correlations of
internal fields at different sites.


One of the fundamental assumptions of the CLS
theory~\cite{ref:CLS,ref:LCS}  is
equipartitioning.  That is,
$p_t({\bf \sigma})$ is assumed to be a constant in a
subspace with a given value of
the single-site spin-field distribution function
$D(\varsigma, h)$ defined in eq.~(\ref{spin-field}).
Our result for $p_t({\bf \sigma})$ shows that this
assumption is violated because a constant $D(\varsigma, h)$ does not mean
a constant value of the field-correlation term $\sum J_{ij}h_{i}h_{j}$.
This term represents correlations of $h_i$ at different sites and hence
cannot be expressed using only the single-site spin-field
distribution function
$D(\varsigma, h)$.
Therefore the CLS theory is not exact.

For a quantitative estimation of various terms in eq.~(\ref{result}),
we have plotted the coefficients $a(t)$ to $c_3(t)$ in Fig.~\ref{fig.1}
for the initial condition $\beta_0=0$.
%
This figure shows that the coefficient $c_3(t)$ of the
field-correlation term is not necessarily small
compared to the others in some time regions, particularly
around $t=2$.  Nevertheless,
in many cases, the
effects of the field-correlation term may not be apparent in physical
observables written only in terms of $D(\varsigma, h)$, such as the
internal energy and the magnetization.  This would be one reason why the
CLS theory predicts the time dependence of these physical quantities
quite accurately, if not exactly, as shown numerically by CLS themselves.

An interesting feature of our formula (\ref{result}) is that, for
small $\beta$ and $\beta_0$,
$p_t({\bf \sigma})$ is well approximated only by the
first-order contribution
given explicitly in eq.~(\ref{1st}).
The first-order term has the form
of a Boltzmann factor
$\exp \{ -\beta_{\rm eff}(t)H({\bf \sigma})\}$.
Thus the system can be regarded as being in equilibrium with effective
temperature $T_{\rm eff}(t)=1/\beta_{\rm eff}(t)$ at any given $t$.
This allows us to calculate various macroscopic physical quantities
in nonequilibrium situations using equilibrium statistical mechanics.
An example is given in Fig.~\ref{fig.2}
for the time development of the internal energy in the case with
$T_0/J=5$ and $T/J=10$.  The center of distribution of the
exchange interactions is $J_0=0$.
%

Concering the generality of the present method,
although the SK model has been discussed explicitly here, it is possible
to apply the same technique to any models including non-random systems
and short-range models. The resulting form of the microscopic probability
distribution is almost independent of details of the model.  Only minor
changes in coefficients are sufficient in many cases.
In particular, in the case of the Hopfield model, the
appearance of various complicated terms in the probablity distribution
(\ref{result}) explains why
the simple two-parameter
dynamics of CS~\cite{ref:CS1,ref:CS2} is not exact as found
numerically.~\cite{ref:ON}

One of the authors (H.N.) thanks Dr.  Hiroshi Takano for useful discussions.

\begin{figure}[hb]
\caption{Time evolution of the coefficients of the high-temperature
expansion. The initial condition is $\beta_0=0$. We set $J=1$.}
\label{fig.1}
\end{figure}

\begin{figure}[hb]
\caption{Time evolution of the internal energy for the initial and
final temperatures $T_0/J=5$ and $T/J=10$, respectively.
The center of 
of exchange interactions is $J_0=0$. The dotted line
represents the asymptotic
value of the energy, $E/NJ=-0.05$.}
\label{fig.2}
\end{figure}

\begin{flushleft}
{\large
 For figures, access the following location
(not the e-mail archive):

 http://www.stat.phys.titech.ac.jp/hp/nishi/PSfigures/

or send your request to nishi@stat.phys.titech.ac.jp.
}
\end{flushleft}

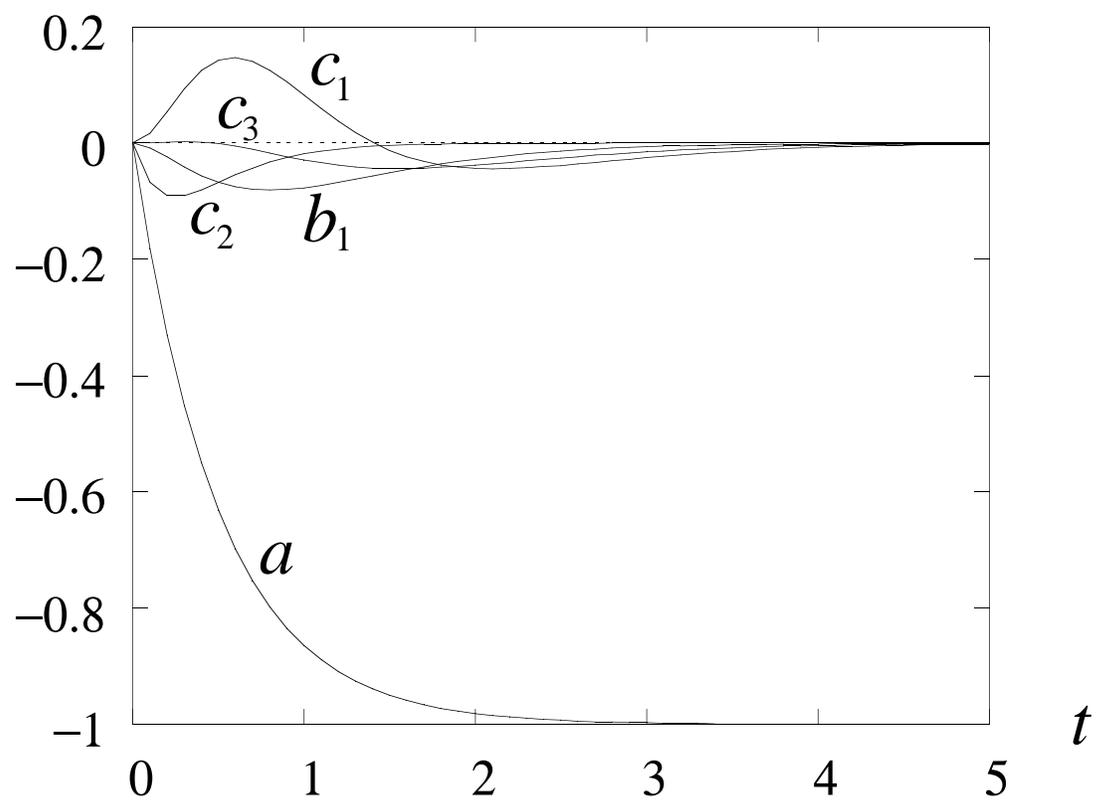
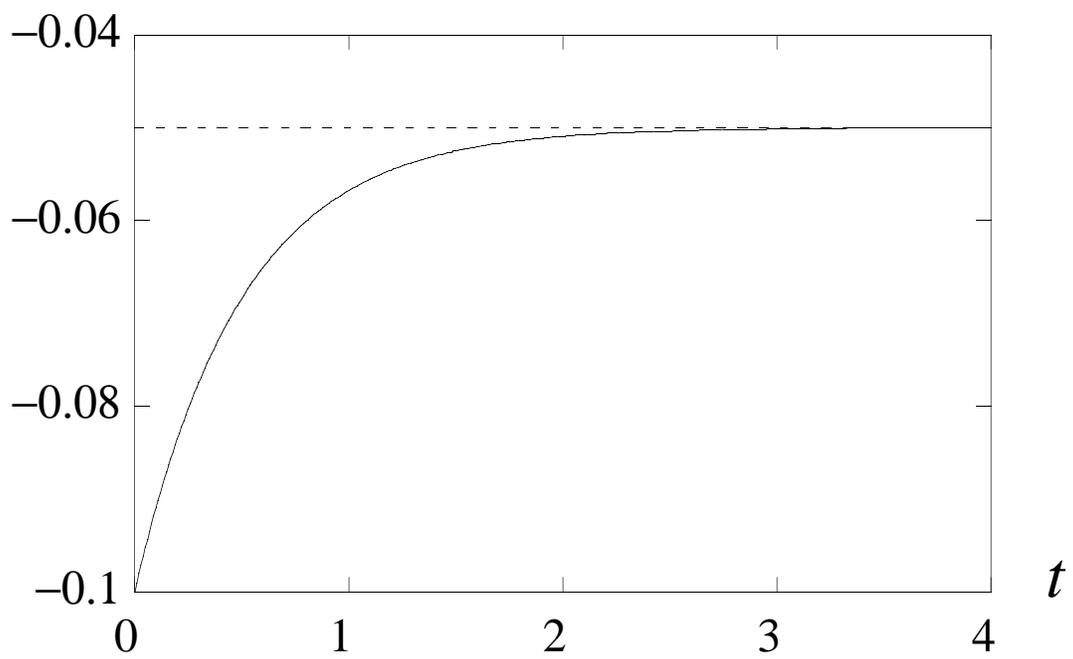
\end{document}